\documentclass[journal]{IEEEtran}
\usepackage[utf8]{inputenc}
\usepackage[english]{babel}
\usepackage{multicol} 
\usepackage[a4paper, margin = 0.6in]{geometry} 

\usepackage{mathtools} 
\usepackage{amsmath}
\usepackage{graphicx} 
\usepackage{float}
\usepackage{fancyhdr} 
\usepackage{url}

\usepackage{hyperref}
\usepackage{comment}
\usepackage[table]{xcolor}
\usepackage{multirow}

\usepackage{caption}
\usepackage{subcaption}

\usepackage{algorithm}
\usepackage{algorithmic}
\usepackage[acronyms]{glossaries}

\usepackage{csquotes}
\usepackage{cleveref}
\usepackage{makecell}
\usepackage{multicol}

\renewcommand\IEEEkeywordsname{Keywords}

\title{LCOE-based Pricing for DLT-enabled Local Energy Trading Platforms} 
\author{
\IEEEauthorblockN{Marthe Fogstad Dynge, Ugur Halden, Gro Klæboe, Umit Cali} \\
\IEEEauthorblockA{Department of Electric Power Engineering, NTNU\\
Trondheim, Norway\\
\{marthe.f.dynge, ugur.halden, gro.kleboe, umit.cali\}@ntnu.no}
}
\date{March 2022}
\makeglossaries
\newacronym{lem}{LEM}{Local Energy Market}
\newacronym{admm}{ADMM}{Alternating Method of Multipliers}
\newacronym{der}{DER}{Decentralized Energy Resources}
\newacronym{res}{RES}{Renewable Energy Sources}
\newacronym{ict}{ICT}{Information and Communication Technologies}
\newacronym{sdr}{SDR}{Supply-Demand Ratio}
\newacronym{fit}{FiT}{Feed-in-Tariff}
\newacronym{lcoe}{LCOE}{Levelized Cost of Electricity}
\newacronym{pv}{PV}{Photovoltaic}
\newacronym{capex}{CAPEX}{Capital Expenditures}
\newacronym{opex}{OPEX}{Operational Expenditures}
\newacronym{dlt}{DLT}{Distributed Ledger Technology}
\newacronym{p2p}{P2P}{Peer-to-Peer}
\newacronym{ev}{EV}{Electric Vehicle}

\begin{document}
\maketitle

\begin{abstract}
Support schemes like the Feed-in-Tariff (FiT) have for many years been an important driver for the deployment of distributed energy resources, and the transition from consumerism to prosumerism. This democratization and decarbonization of the energy system has led to both challenges and opportunities for the system operators, paving the way for emerging concepts like local energy markets. The FiT approach has often been assumed as the lower economic bound for a prosumer's willingness to participate in such markets but is now being phased out in several countries. In this paper, a new pricing mechanism based on the Levelized Cost of Electricity is proposed, with the intention of securing profitability for the prosumers, as well as creating a transparent and fair price for all market participants. The mechanism is designed to function on a Distributed Ledger Technology-based platform and is further set up from a holistic perspective, defining the market framework as interactions in a Cyber-Physical-Social-System. Schemes based on both fixed and variable contracts with the wholesale supplier are analyzed and compared with both the conventional FiT and to its proposed replacement options. The results show a cost reduction for the consumers and a slight loss in revenue for the prosumers compared to the FiT scheme. Comparing it to the actual suggested replacements to the FiT, however, it is clear that the pricing mechanism proposed in this study provides a substantial increase of benefits for both prosumers and consumers.
\end{abstract}

\begin{IEEEkeywords}\raggedright
Cyber-Physical-Social-Systems, Distributed Ledger Technology, Electricity Pricing, Levelized Cost of Electricity, Local Energy Markets 
\end{IEEEkeywords}

\section{Introduction}
Along with the pressing need to mitigate climate change, the world's energy sector is undergoing a remarkable transition. To fulfill the goal of decarbonization, as well as reaching the United Nations Sustainable Development Goal (UN SDG) \#7 (access to affordable, reliable, sustainable, and modern energy for all), renewable energy sources are rapidly entering the system. The affordability of small-scale, \glspl{der} is thus increasing correspondingly. With the simultaneous maturing of Information and Communication Technologies, such as \gls{dlt}, previously passive consumers are able to take a more active role in the energy system, transitioning to being prosumers. This can further accelerate the enrollment of renewable energy and unlock flexibility in the distribution system for future grid planning and operation. However, the introduction of prosumers into the distribution system may lead to challenges in terms of operational stability and security, as well as social and legislative issues. To both enhance the benefits of the democratization of the energy system and to control the challenges, \glspl{lem} have emerged as a promising concept \cite{Tushar2021Peer-to-peerChallenges,Bjarghov2021DevelopmentsReview}. 

One of the key determinants of a prosumer's willingness to participate in a \gls{lem} is the community trading price \cite{Hahnel2020BecomingCommunities}. Multiple studies have proposed different strategies for both bidding processes and the actual price setting, all including different aspects to be considered in a fair market price. In general, one can categorize the different schemes proposed in the literature into auction-based \cite{Lin2019ComparativeMarkets,Ghorani2018OptimalMarkets,Wang2020ASystem,Vieira2021Peer-to-peerContracts}, optimization-based \cite{Morstyn2020IntegratingPricing,Paudel2019PricingApproach,Ullah2020DistributedGrid,Morstyn2019MulticlassPreferences,Lezama2020LearningOptimization}, game theory-based \cite{Jiang2020ElectricityEnvironment,Fernandez2021APeers} and cost-sharing \cite{Cali2019EnergyMarkets,Grzanic2021ElectricityUncertainty,Long2018Peer-to-peerMicrogrid} methods.

Auction-based methods rely on the active participation of each market participant by placing bids and offers to the market platform. Dynamic price adjustments are analyzed in \cite{Wang2020ASystem} and show advantages in terms of higher social welfare and reduced environmental impacts compared to other bidding strategies. Two auction mechanisms, Discriminatory and Uniform k-mean, are analyzed in \cite{Lin2019ComparativeMarkets}, as well as the impact of different bidding strategies. The analysis suggests that the discriminatory approach outperforms the uniform mechanism in terms of market participation, but is more sensitive to market conditions and can lead to more fluctuating trading prices. Employing a game-theoretic bidding strategy where the participants compete to bid for the best price obtains a near-optimal economic efficiency according to this analysis.    

Although game-theoretic approaches have gained popularity in the literature, market participants are still required to take an active role in price-setting games. A common approach in modeling the \gls{lem}
interactions are Stackelberg games, where the prosumers act as leaders and the consumers act as followers, as done in \cite{Jiang2020ElectricityEnvironment} and \cite{Fernandez2021APeers}. A similar game is analyzed in \cite{Anoh2020}, showing higher social welfare among the market participants compared to direct \gls{p2p} trading. A socially optimal solution, which is also Pareto optimal, is obtained through a cake-cutting game in \cite{Tushar2017PriceApproach}, with discriminate pricing. However, both auction-based and game theoretic market structures are high-demanding strategies in terms of participation and may seem intimidating to some agents. This can in turn discourage the establishment of \glspl{lem}.

Other pricing mechanisms use different optimization techniques to set the \gls{lem} prices implicitly, through dual prices of an optimal power flow model \cite{Morstyn2020IntegratingPricing} or a general market clearing problem \cite{Ullah2020DistributedGrid}. Others propose more direct price-setting optimization algorithms, like consensus-based \gls{admm} \cite{Paudel2019PricingApproach}. In \cite{Morstyn2019MulticlassPreferences}, the \gls{lem} price is calculated based on a multi-class energy clearing problem, solved through the \gls{admm} approach, considering the different preferences of the market participants. 

In most cost-sharing methods, the community costs are distributed ex-post, and the final price is set based on different balancing criteria. The prosumers are thus considered as price-takers and cannot actively influence the \gls{lem} price. In \cite{Cali2019EnergyMarkets}, a \gls{sdr} is calculated based on the energy balance within the community at each time step. The local trading price is linearly set between an upper and lower bound, based on this ratio. Two incentive extensions to this method are also proposed. A similar method combined with a compensation rate is analyzed by \cite{Long2018Peer-to-peerMicrogrid} and prosumers' preferred level of participation is taken into account in the approach proposed by \cite{Liu2017Energy-SharingProsumers}. The \gls{sdr} approach is compared with the Bill Sharing Scheme and the Mid-Market Rate in \cite{Grzanic2021ElectricityUncertainty}, showing the disadvantages of the existing cost-sharing schemes, while proposing an improved, two-stage \gls{sdr} mechanism. The same three methods are compared in \cite{Zhou2018EvaluationFramework}, illustrating an outperformance of the \gls{sdr} method compared to the other two in terms of the overall performance.

In this paper, the \gls{sdr} pricing mechanism is further explored. With the \gls{fit} schemes being phased out in, e.g., Germany, a new limit on the minimum trading price based on the \glspl{der}' \gls{lcoe} is proposed to ensure both prosumer and consumer profitability. The usage of \gls{lcoe} as a pricing instrument for \glspl{lem} is limited within the current literature. In \cite{Vieira2021Peer-to-peerContracts}, it is used together with the national grid price to form a truncated normal distribution as a basis for two different auction mechanisms. It is also used as a minimum trading price for a price matching algorithm in \cite{An2020DeterminingMicrogrid}. However, further research is required and thus the contributions of this paper are as follows:
\begin{itemize}    
    \item Rethinking the existing convention of the FiT as the lower bound of prosumers' willingness to participate in \glspl{lem}, an LCOE-based mechanism is proposed to sustain the viability of the prosumers' investments in production capacity.
    \item A fair and transparent pricing mechanism for \glspl{lem} is proposed, distributing the costs and benefits of the local energy trading based on each market participant's contribution to the supply-demand balance.
    \item The impacts of the proposed LEM pricing mechanism are investigated using a techno-economic analysis from both prosumers' and consumers' points of view.
    \item Conceptualization of a potential DLT-based local energy trading platform that can use the proposed LEM pricing mechanism as an integral part. 
\end{itemize}

\section{Distributed Ledger Technology in Local Energy Markets}
\gls{dlt} can be defined as a distributed digital consensus and record-keeping system whose main purpose is to eliminate unnecessary third parties and build trust in a generally untrustful environment such as the internet. The main working principle of DLT is to allow the transactions (either of data or actual payments) to be logged, distributed and saved immutably in private, public, or hybrid networks depending on the architecture of the system. Meanwhile, for validation and authentication of these transactions, various consensus systems are utilized, with the Proof-of-Work (PoW) and Proof-of-Stake (PoS) being the most dominant ones \cite{Andoni2019BlockchainOpportunities}. There are various use cases within the applications of DLT in the energy sector. However, to keep this section concise, only peer-to-peer energy trading, electric vehicle charging, and payment settlement parts will be explained further as the core use cases of LEMs.

As the number of smart meters, DER, and various two-way communication equipment entering the power system keep increasing, a paradigm shift in the energy sector is inevitable. To enable this transition towards energy democratization, DLT can be utilized to remove the need for Trusted Third Parties (TTP), provide privacy towards market participants, and speed up the transaction validation times. This has further opened the possibility to form peer-to-peer markets, where smaller agents can trade energy directly between each other \cite{Andoni2019BlockchainOpportunities}. Several projects, albeit in the pilot phase, have developed working trading platforms with DLT and various tokens to enable this kind of direct trading \cite{Khorasany2021LightweightTrading}\footnote{IEEE Spectrum. (2018). Startup Profile: ME SOLshare’s “Swarm Electrification” Powers Villages in Bangladesh. \url{https://spectrum.ieee.org/startup-profile-me-solshares-swarm-electrification-powers-villages-in-bangladesh} [Accessed February 13, 2022].}. However, most existing projects suffer from scalability issues both in terms of DLT and physical infrastructure integration. Thus, more research needs to be performed for utilization in real-world applications. 

Due to the ongoing electrification of transport and the increasing popularity of electric vehicles, the electricity and transport sectors are becoming more entangled. However, for widespread adaptation and utilization, significant barriers still exist such as unavailability or scarcity of public charging infrastructures. Additionally, the management of high-resolution data and complex transactions that need to happen and be validated as quickly as possible poses another barrier. DLT-enabled LEM networks can be utilized in this step as a charging regulator, which can perform the required power and token transactions in a fast, secure, and private way \footnote{WePower. (2022). WePower: Secure energy buyers faster and cheaper. \url{https://wepower.com/project-owners.html} [Accessed February 13, 2022].}\footnote{Besnainou, J. (2018). Autonomous datasets and v2x transactions: Blockchain in mobility pilots getting traction. \url{https://www.cleantech.com/autonomous-datasets-and-v2x-transactions-blockchain-in-mobility-pilots-getting-traction/} [Accessed February 13, 2022].}.
\section{Methodology}\label{sec:method}
\begin{figure}[h]
    \centering
    \includegraphics[width=3.5in]{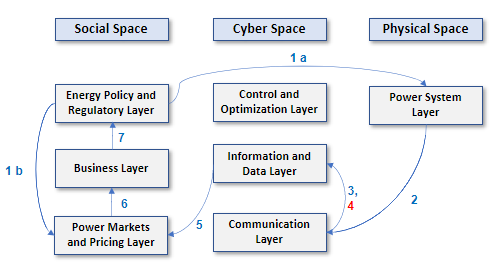}
    \caption{Energy CPSS Interactions of the system.}
    \label{fig:CPSS}
\end{figure}

The modern energy system can be classified as a combination of cyber, physical, and social subsystems, creating a cyber-physical-social-system (CPSS), as described in \cite{Cali2021NovelProblem}. The overall market framework of this paper can be explained through the interaction between these layers, illustrated in Fig.~\ref{fig:CPSS}.     

The main underlying assumption of the \gls{lem} is that necessary energy policies and regulations are in place. In that case, private households are not only allowed but encouraged through legal provisions to form transactive energy markets with their neighbors. The policies and regulations will thus influence the deployment of new physical components of the power system, such as DLT-enabled electric meters, that will be operated in the \gls{lem} and support the proposed approach (\textit{Process (1a)}). They will also affect how the \gls{lem} is structured and how the local energy price will be set and communicated with the \gls{lem} participants (\textit{Process (1b)}). The interaction between the CPSS layers in the operational phase is designed as follows.

\textit{Process (2)}. The \gls{lem} aggregator facilitates the routing of energy to satisfy the needs and preferences of the market participants, without violating the physical boundaries of the system. Smart meters provide information about generation and consumption for each market participant and communicate this information to the \gls{lem} platform. 

\textit{Process(3)}. The information from the neighborhood smart meters is collected and processed by the \gls{lem} platform.

\textit{Process (4)}. DLT is used to ensure trust is built between transacting parties, generate immutable track recording, and secure sharing of information and transactions on the market platform. The fourth process of the framework is marked red to symbolize that it is not implemented in this paper's analysis but is assumed to be in place for the real-life implementation of the \gls{lem} platform where DLT can be deployed.

\textit{Process (5)}. The processed data regarding the market participants' generation and consumption profiles are retrieved by the market mechanism. The \gls{lem} price for each timestamp is then obtained. Pricing information is then sent back to the DLT network.  Since this information is sensitive and not expected to overload the DLT network, such data can be stored in an on-chain database. 

\textit{Process (6)}. Based on the pricing results, techno-economic analyses are made to investigate the economic conditions of the neighborhood, as well as the individual gains or losses as a result of participating in the \gls{lem}.

\textit{Process (7)}. The experience from the techno-economic evaluation may form new suggestions for policy and regulation improvements.

\begin{figure*}[ht]
    \centering
    \includegraphics[width=\textwidth]{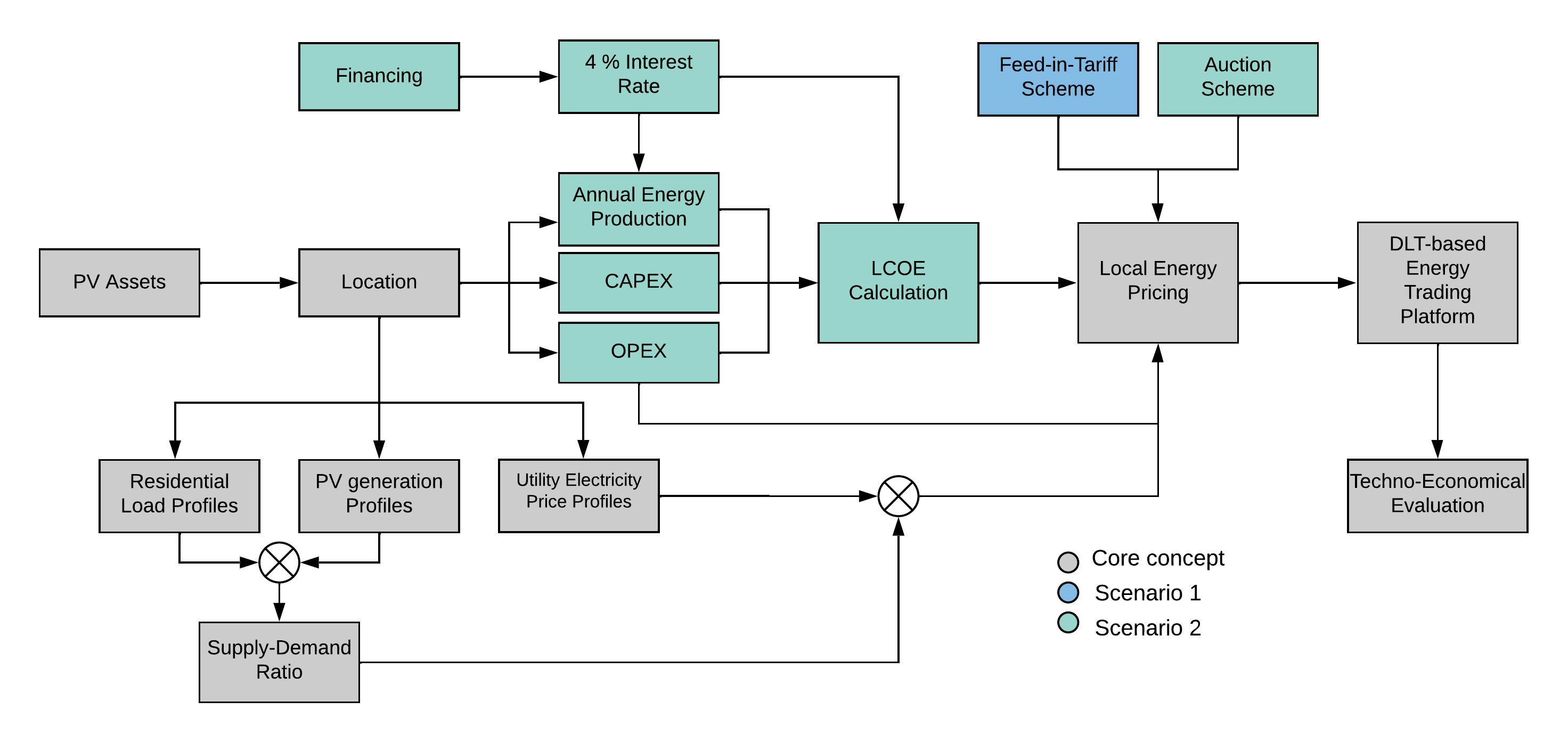}
    \caption{Flowchart of overall pricing mechanism.}
    \label{fig:methodology}
\end{figure*}

\subsection{Replacing the Feed-in-Tariffs}
Government-funded \gls{fit} schemes have for decades ensured renewable energy producers, from utility-scale to small households, a steady pay-back for their energy surplus fed back to the grid. In Germany alone, the generous scheme has contributed to the increasing share of renewable power production, from 3.5\% in 1990 to 40.9\% in 2021 \footnote{AGEB - AG Energiebilanzen e.V. (2021). Electricity generation by energy sources (electricity mix) from 1990 to 2021 (in TWh) Germany as a whole (in German). \url{https://ag-energiebilanzen.de/wp-content/uploads/2021/02/Strommix-Dezember2021.pdf} [Accessed March 8, 2022]}. However, as the 20-year guarantee for the pioneer installations is reaching its end, the German policymakers are intending to phase out the FiTs\footnote{Appunn, K. and Wehrmann, B. (2019). Germany 2021: when fixed feed-in tariffs end, how will renewables fare? - Energy Post. \url{https://energypost.eu/germany-2021-when-fixed-feed-in-tariffs-end-how-will-renewables-fare/} [Accessed February 7, 2022]}. The business model replacing this scheme will be formed by 2027, and a more market-driven structure is expected. New larger scale installations have for recent years switched to auctions, where the price is set by the bidders and the maximum volume built is determined by the government. Although installations below 750 kW are still excepted from participating in the auction processes, it is expected that the remuneration price offered to prosumers, if any, for their feed-in power will stabilize around the same level as the auction prices.

\subsection{Levelized Cost of Electricity}
To provide incentives for both consumers and prosumers to participate in a \gls{lem}, a general assumption is that the consumer should pay a lower price for electricity than from the wholesale market, and the prosumer should get a higher price than from any other undertaker. Thus, a stable \gls{lem} could only exist if the \gls{lem} price, $p_t$, stays within the limits of the highest price a prosumer can get, $p_L$, and the lowest price a consumer can get, $p_U$, outside the \gls{lem}:

\begin{equation}
    p_L \leq p_t \leq p_U
\end{equation}

Based on the current policies in most electricity markets, $p_L$ can generally be defined as the \gls{fit}. However, as Germany and other countries are phasing out the \gls{fit} scheme, another more long-term lower limit must be developed.

With no specific schemes for the retail market to procure excess energy from the neighborhood, another priority for the prosumers will be to recover their investment costs for energy resources. Thus, the \gls{lcoe} is considered to be a logical new lower boundary for the \gls{lem} price as it reflects the average cost of producing power from the \gls{pv} system over its lifetime. 

\begin{equation}
    LCOE = \dfrac{I_0 + \sum^N_{n=1}\dfrac{A_n}{(1+R)^n}}{\sum^N_{n=1}\dfrac{M_{n,el}}{(1+R)^n}}
    \label{eq:lcoe_new}
\end{equation}

The main components of the \gls{lcoe} are the fixed and variable costs related to investing in a \gls{pv} system, divided by the total amount of electricity produced over the system's lifetime, as represented in Eq.~\eqref{eq:lcoe_new}. $I_0$ denotes the initial investment expenditures, $A_n$ is the fixed and variable costs for the year $n$ and $M_{n,el}$ is the total electricity generated by the system in year $n$. $R$ represents the weighted average cost of capital.  The \gls{capex} related to the initial investment costs of the system is defined by Eq.~\eqref{eq:capex}. 

\begin{equation}
    CAPEX = PV_{cap} \cdot (C_{E}+C_{Dl}+C_{Il}+C_{P}+C_{O})
    \label{eq:capex}
\end{equation}

Here, $C_E$ denotes the equipment costs per kWp, $C_{Dl}$ and $C_{Il}$ are the direct and indirect labor costs, whereas $C_p$ represents the project permission costs and $C_O$ the overhead costs such as project administration. $PV_{cap}$ is the installed PV capacity.

The variable costs are represented by the \gls{opex}. These are costs related to the annual operation and maintenance throughout the lifetime of the system. In general, the OPEX account for a much smaller percentage than the CAPEX \cite{Vartiainen2020}. The LCOE value is naturally individual for each of the \gls{lem} prosumers. The LCOE value referred to further in this study is thus assumed to be the average across the market participants, for simplicity. 

\subsection{Supply-Demand Ratio}
The proposed pricing mechanism in the following sections is a modification and improvement of the method introduced by \cite{Cali2019EnergyMarkets}. The overall concept is to let the ratio between supply and demand in the \gls{lem} determine the local trading price, and thus distribute the costs in accordance with each market participant's contribution to the \gls{lem}. 

The total surplus in the neighborhood, and thus the amount available for local trade, is calculated based on each market participant's net surplus after ensuring its self-consumption. Hence, the surplus is defined as the difference between the amount generated by agent $i$ in timestep $t$, $g_{it}$, and the load of the same agent in said timestep $l_{it}$, as described in \eqref{eq:surplus}. 

\begin{equation} s_{it}= \begin{cases} g_{it}-l_{it},&\text {if} ~g_{it}>l_{it} \\ 0,&\text {if} ~g_{it}\leq l_{it}. \end{cases}\label{eq:surplus}\end{equation}

The net demand of the neighborhood is similarly defined in \eqref{eq:demand} as the net demand of agent $i$ in timestep $t$ after consuming its self-generated power. 

\begin{equation} d_{it}= \begin{cases} l_{it}-g_{it},&\text {if} ~l_{it}>g_{it} \\ 0,&\text {if} ~l_{it}\leq g_{it}. \end{cases}\label{eq:demand}\end{equation}

The \gls{sdr} further used as a base for the local energy trading price is then defined in \eqref{eq:ratio} as the ratio between the total surplus and demand in the neighborhood. 
\begin{equation} r_{t}= \frac {\sum \limits _{i\in P_{t}}^{} s_{it}} {\sum \limits _{i\in C_{t}}^{} d_{it}}\label{eq:ratio}\end{equation}

\subsection{Pricing Scheme}
The overall pricing mechanism proposed in this study is described by the flowchart in Fig.~\ref{fig:methodology}. The grey boxes represent the steps involved in the core concept, meaning that they are involved regardless of the different variations later described. The green and blue boxes represent the scenario-dependent components, related to the scenarios described in Section \ref{sec:case}. In addition to this, two variations within the ''Local Energy Pricing'' box are proposed in this study and will be further explained in the following subsections.

\subsubsection{Fixed Upper and Lower Price Thresholds} \label{sec:method_fixed}
In markets where prosumers are offered a fixed price for their surplus and most consumers are assigned to fixed-price contracts, a stable \gls{lem} situation can be obtained if 

\begin{equation}
   p_L \leq LCOE \leq p_t \leq p_U
   \label{eq:balance}.
\end{equation}

Under this criterion, there will be incentives for both consumers and prosumers to participate in the \gls{lem} as well as investing in new generation capacity. 

When there is no surplus in the market, hence $r_t = 0$, all remaining demand must be covered by purchasing from the wholesale market. Thus, the \gls{lem} price equals its upper limit, $p_t = p_U$, the wholesale energy price.

With $0 < r_t < 1$, the demand in the neighborhood can be partially covered by local surplus. When there is a limited surplus, the \gls{lem} price tends towards the upper limit $p_U$.

\begin{equation} \lim _{r_{t}\to 0^{+}} p_{t} =p_{U}\label{eq:upper_limit}\end{equation}

On the other end, when the local surplus almost covers the entire demand, the \gls{lem} price orients towards the lower selling limit. By the balancing criterion in \eqref{eq:balance}, this should equal the LCOE when $r_t<1$.

\begin{equation} \lim _{r_{t}\to 1^{-}} p_{t} =LCOE.\label{eq:lower_limit}\end{equation}

On sunny days with low demand, situations can occur where the surplus exceeds the demand in the neighborhood, hence $r_t >= 1$. In these cases, the abundant power will be sold to the utility and the net consumers in the neighborhood can procure power locally for the same price. Thus, the \gls{lem} price equals $p_t = p_L$.

The final \gls{lem} price could thus be defined by a convex combination of the LCOE and $p_U$ as long as the market is not fully saturated, as well as collapsing to $p_L$ when the \gls{lem} is saturated.

\begin{equation} p_{t}= \begin{cases}r_t \cdot LCOE + (1-r_t)\cdot p_U,&\text {if} \quad r_t < 1 \\ p_L,&\text {if} \quad r_t \geq 1. \end{cases}\label{eq:price}\end{equation}  

\subsection{Dynamic Upper and Fixed Lower Price Thresholds}
\label{sec:method_dyn}
In some areas, it is more common for end-users to choose variable pricing contracts. This variable price can be based on the wholesale electricity market price, where the consumer pays a markup to buy power from the grid from a supplier. Similarly, the wholesaler deducts a margin from the spot price to make a margin on the marketing of local production. We can then define a time-variable upper threshold for the \gls{lem} as follows:

\begin{equation}
     p_{Ut} = \alpha_t + \mu_{cons} 
\end{equation}

where  $\mu_{cons}$ is the mark-up of the wholesaler and $\alpha_t$ is the time-dependent wholesale electricity price. The lower threshold, $p_L$ is fixed and is here defined as the price a prosumer would get if selling its surplus to the utility.

With dynamic pricing thresholds and as \glspl{der} struggles to reach total grid parity in many countries, situations may occur where $p_{Ut}<$ LCOE. Provided that $p_{Ut}>0$, the prosumer would still prefer to sell for this price within the \gls{lem}, as it would be more profitable than curtailing without any storage options. If $p_{Ut}<0$, it is assumed that the prosumers can curtail their excess production, rather than paying the utility. While LCOE $< p_{Ut}$ the \gls{lem} price follows the same principles as defined in Section~\ref{sec:method_fixed}, thus resulting in the final pricing scheme illustrated by Algorithm~\ref{alg:dynamic}.    

\begin{algorithm}
\caption{Algorithm for \gls{lem} pricing scheme with dynamic thresholds.}
\label{alg:dynamic}
\begin{algorithmic}
\FOR{each $t$ in $T$}
\STATE Calculate $q_t$
\IF{$p_{Ut} <$  LCOE}
\STATE $p_t \gets p_{Ut}$
\ELSE
\IF{$r_t = 0$}
\STATE $p_t \gets p_{Ut}$
\ELSIF{$r_t<1$}
\STATE $p_t \gets r_t \cdot$ LCOE $+ (1-r_t)\cdot p_{Ut}$
\ELSIF{$r_t >= 1$}
\STATE $p_t \gets p_{L}$
\ENDIF
\ENDIF
\ENDFOR
\end{algorithmic}
\end{algorithm}

\subsection{Cost-Benefit Analysis}
To evaluate the advantages of the proposed pricing mechanisms, a cost-benefit analysis of both the total market costs and the individual benefits was conducted. As with the \gls{lem} price itself, the costs are distributed differently depending on the \gls{sdr}, $r_t$. When $0<r_t\leq 1$, all surplus energy is traded within the \gls{lem}, and the consumers cover the remaining demand by procuring from the utility. Hence, the total neighborhood cost for each time step is defined as

\begin{equation} c_{t}=\left ({\sum _{i\in P_{t}} s_{it} }\right)\cdot p_{t}+\left ({\sum _{i\in C_{t}} d_{it} -\sum _{i\in P_{t}} s_{it} }\right)\cdot p_{Ut}.
\end{equation}

Here, $P_t$ is defined as the set of prosumers in timestep $t$ in the \gls{lem}, whereas $C_t$ is the set of consumers in timestep $t$. If the \gls{lem} is saturated, i.e., $r_t>1$, all electricity demand in the neighborhood would be covered by locally traded energy, sold by the local trading price, $p_t$.
\begin{equation}
    c_t = \sum _{i\in C_{t}} d_{it}\cdot p_t
\end{equation}

This total cost is distributed among the market participants based on their share of the total energy requirement in the neighborhood.
 
\begin{equation} c_{it}=\frac {d_{it}}{\sum _{i\in C_{t}} d_{it}}\cdot c_{t}\end{equation}

Applying the definition of $r_t$ and $c_t$, this can be reformulated to

\begin{equation} c_{it}=d_{it}\cdot \left ({r_{t}\cdot p_{t}+ (1-r_{t})\cdot p_{Ut} }\right).\end{equation}

Both the total cost $c_t$ and the individual cost $c_{it}$ are defined as pure procurement costs and do not account for the prosumers' revenue for selling their surplus. This revenue is distributed post-trade depending on each prosumer's contribution to the supply-demand balance in the \gls{lem} at each time step and the \gls{lem} trading price at that hour. When $r_t>0$, all surplus is sold through the \gls{lem} for the local price. 
\begin{equation} y_{t}=\sum _{i\in P_{t}} s_{it}\cdot p_{t} \end{equation}

The individual prosumer's revenue is distributed post-trade depending on its contribution to the total surplus in the \gls{lem}, in the same manner as the procurement costs. 

\begin{equation}
    y_{it}=\frac {s_{it}}{\sum _{i\in P_{t}} s_{it}}\cdot y_{t} \label{eq:ind_profit}
\end{equation}

By using the definition of both $r_t$ and $y_t$, \eqref{eq:ind_profit} can be reformulated to

\begin{equation}
    y_{it} = s_{it} \cdot \left(\dfrac{1}{r_t}\cdot p_t + \Big(1-\dfrac{1}{r_t}\Big)\cdot p_{L} \right).
\end{equation}

The net cost of the prosumers is subsequently defined as

\begin{equation}
    c^{net}_{it} = c_{it} - y_{it}.
\end{equation}

\section{Local Electricity Market Structure and Analyzed Scenarios} \label{sec:case}
\begin{figure}[ht]
    \centering
    \includegraphics[width=3.5in]{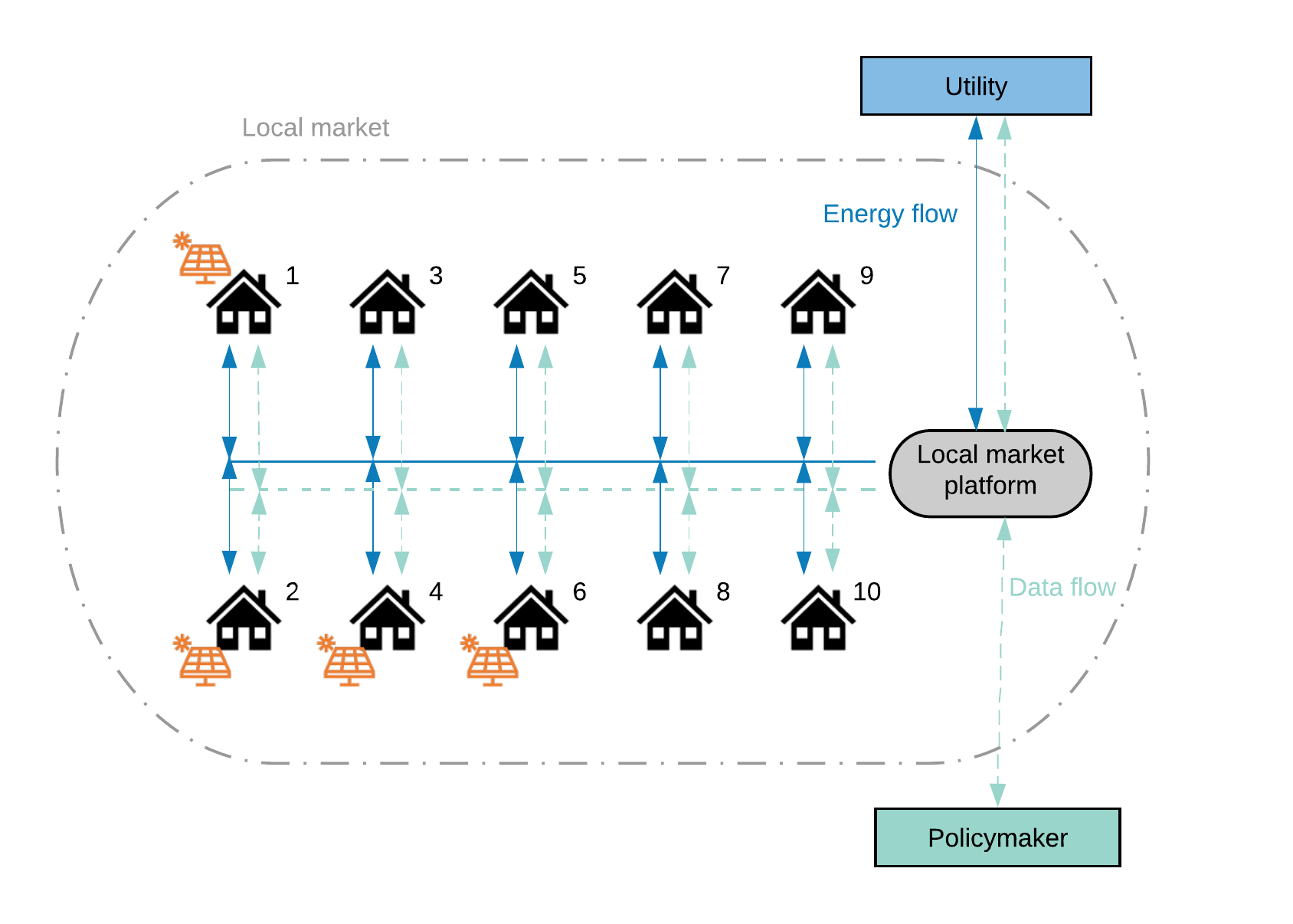}
    \caption{Illustration of the analyzed local market.}
    \label{fig:neighborhood}
\end{figure}

To analyze the effects of the pricing strategies proposed in this paper, a virtual \gls{lem} platform was designed. The market consists of 10 participants, randomly assigned the role of prosumers and consumers. The structure and distribution of \glspl{pv} are illustrated in Fig.~\ref{fig:neighborhood}.  
As illustrated in Fig.~\ref{fig:neighborhood}, both energy and data are exchanged both internally in the \gls{lem} and with entities outside. When the \gls{lem} is saturated, prosumers can sell their excess surplus to the utility through the \gls{lem} platform. From a longer-term perspective, policymakers evolve their policies regarding \glspl{lem} through data collected from the \gls{lem} platform. 

For this particular case study, artificial load profiles with hourly granularity for each of the market participants were generated with the open-source Artificial Load Profile Generator developed by \cite{Hoogsteen2017AManagement}. \gls{pv} production time-series were retrieved through the open-source tool Renewables Ninja \cite{Pfenninger2016Long-termData}, which generates profiles for a given geographical area based on the meteorological database NASA MERRA-2 \cite{Rienecker2011MERRA:Applications}. The coordinates of Freiburg im Breisgau were used to estimate the yearly \gls{pv} production series of a neighborhood in Southern Germany.

Pricing strategies for end-users assigned to fixed price and dynamic price contracts are analyzed in this study. In this case, the fixed utility price is set to 30.46 \texteuro/MWh, according to German levels of 2019 \cite{PriceGermany}. The \gls{lem} price thresholds in the dynamic case are based on the wholesale electricity market price, in this case, set to the spot price of Germany in 2019 \cite{FraunhoferISESpotEnergy-Charts}. The supplier markup for consumers, $\mu_{cons}$, is set to equal 10\% of the spot price. 

The novel proposal of this study is the usage of LCOE in the \gls{lem} price setting. However, for benchmarking purposes, scenarios with the \gls{fit} in the place of the LCOE value for both pricing mechanisms are analyzed. The FiT value is set to 6 \texteuro/MWh following German rates \cite{Renews} in 2019. The LCOE is set to equal 8 \texteuro/MWh, as calculated for German PV systems subject to a 4\% interest rate in \cite{Halden2021DLT-basedInvestments}. This is in accordance with the range between 6 and 11 \texteuro/MWh found by Fraunhofer ISE \cite{Kost2021} for small PV rooftops systems in Germany in 2021. In the cases where LCOE is used, $p_L$ is defined as the average auction value that is offered outside the LEM and is set to equal 5 \texteuro/MWh \cite{Renews}. Following the flowchart in Fig.~\ref{fig:methodology}, Scenario 1 refers to the cases where the \gls{fit} is used, for both fixed and dynamic pricing thresholds. Scenario 2 refers to the usage of LCOE.

\section{Results}
\begin{figure}[h]
    \centering
    \includegraphics[width=3.5in]{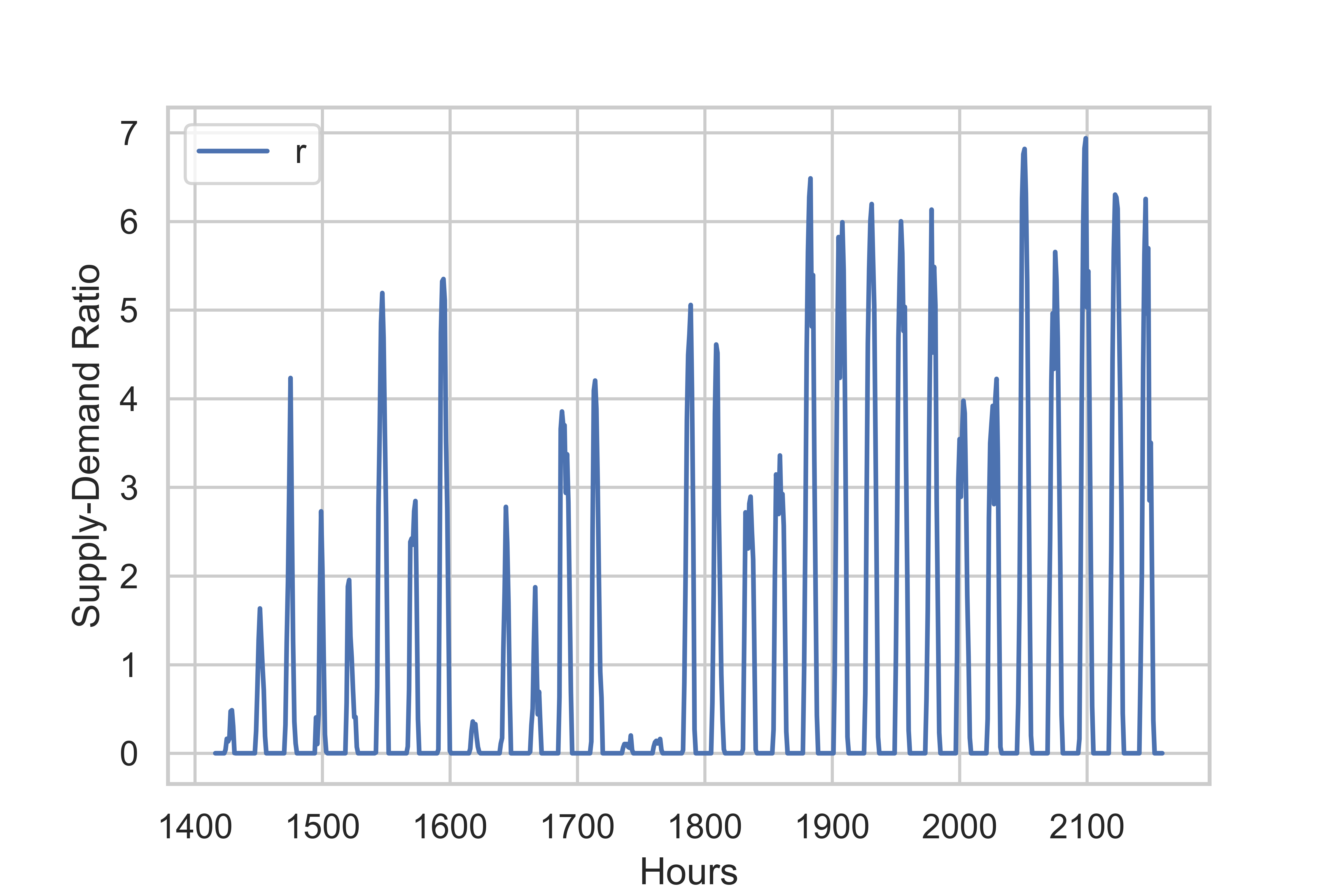}
    \caption{\gls{sdr}, $r_t$, in March.}
    \label{fig:q_march}
\end{figure}

To capture seasonal variations in the \gls{sdr} profile, hourly simulations for an entire year were executed for all cases. To illustrate how the \gls{sdr} varies over time, the hourly values for March are shown in Figure~\ref{fig:q_march}. It is clear that the $r_t > 1$ condition occurs quite frequently, indicating a slight overscaling of the number of prosumers in the neighborhood.

As the scenarios with fixed and dynamic \gls{lem} pricing thresholds are not directly comparable, they will be presented in separate sections.

\subsection{Fixed Pricing Thresholds}
\begin{figure}[h]
    \centering
    \includegraphics[width=3.5in]{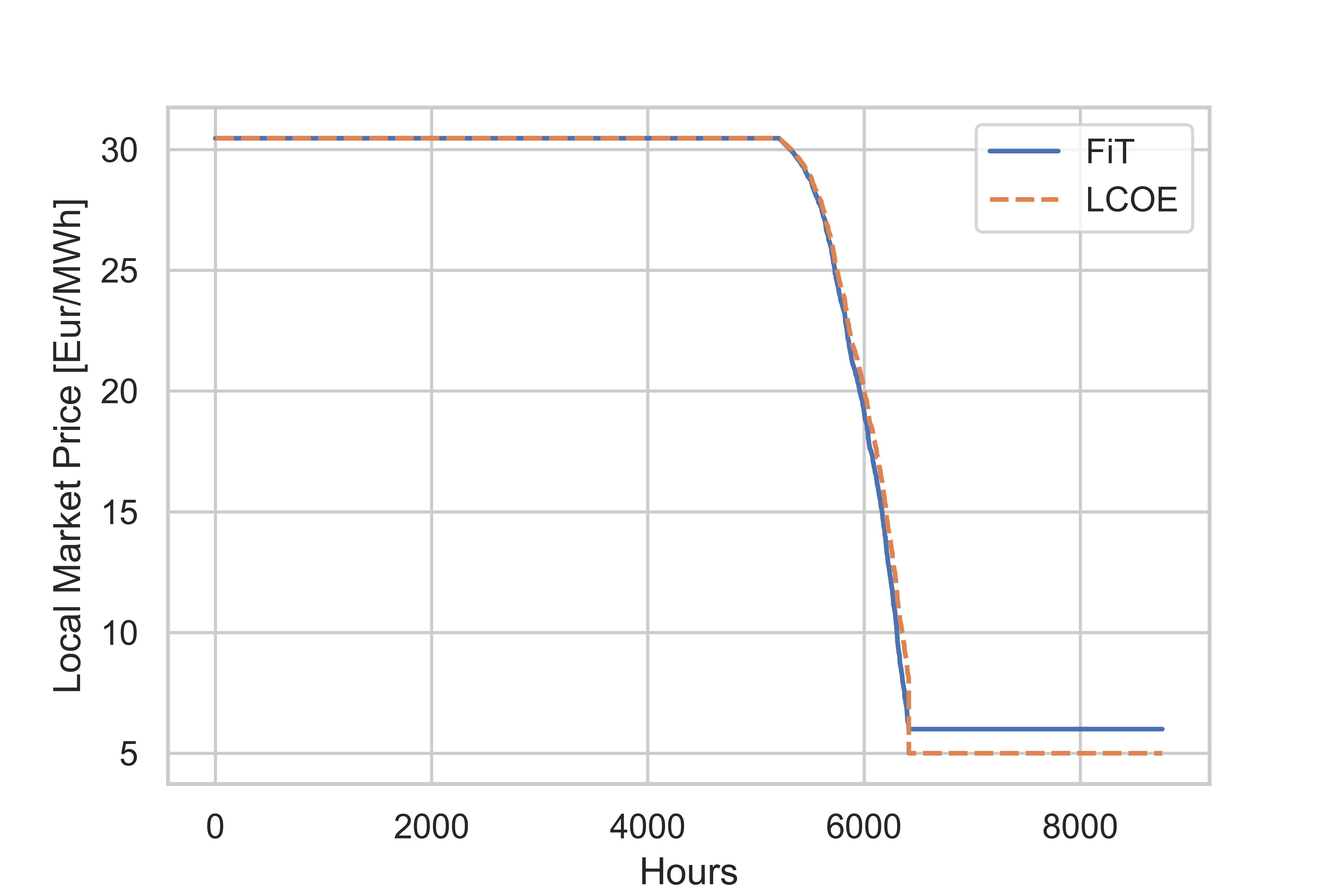}
    \caption{Duration curves of the local energy market prices for the scenarios with fixed utility price.}
    \label{fig:Duration_comp_static}
\end{figure}

Firstly, the scenarios with fixed pricing thresholds are analyzed. The resulting \gls{lem} prices for each hour of the simulated year are sorted in descending order and plotted for the two different scenarios to form a duration curve in Figure~\ref{fig:Duration_comp_static}. Based on the curve's characteristics, it is clear that the \gls{lem} is operative for almost half of the year. As the LCOE is slightly higher than the \gls{fit}, the \gls{lem} price is marginally higher while $r_t < 1$. When the market is saturated, the \gls{lem} price collapses to either equal the \gls{fit} or the auction price, and thus the prices in the LCOE-based scenario are lower. The declining curve between the upper and lower bound is quite steep, indicating that the market is mostly saturated when it is active, in accordance with the \gls{sdr} levels in the sample period in March in Figure~\ref{fig:q_march}. 

\begin{figure}[h]
    \centering
    \includegraphics[width=3.5in]{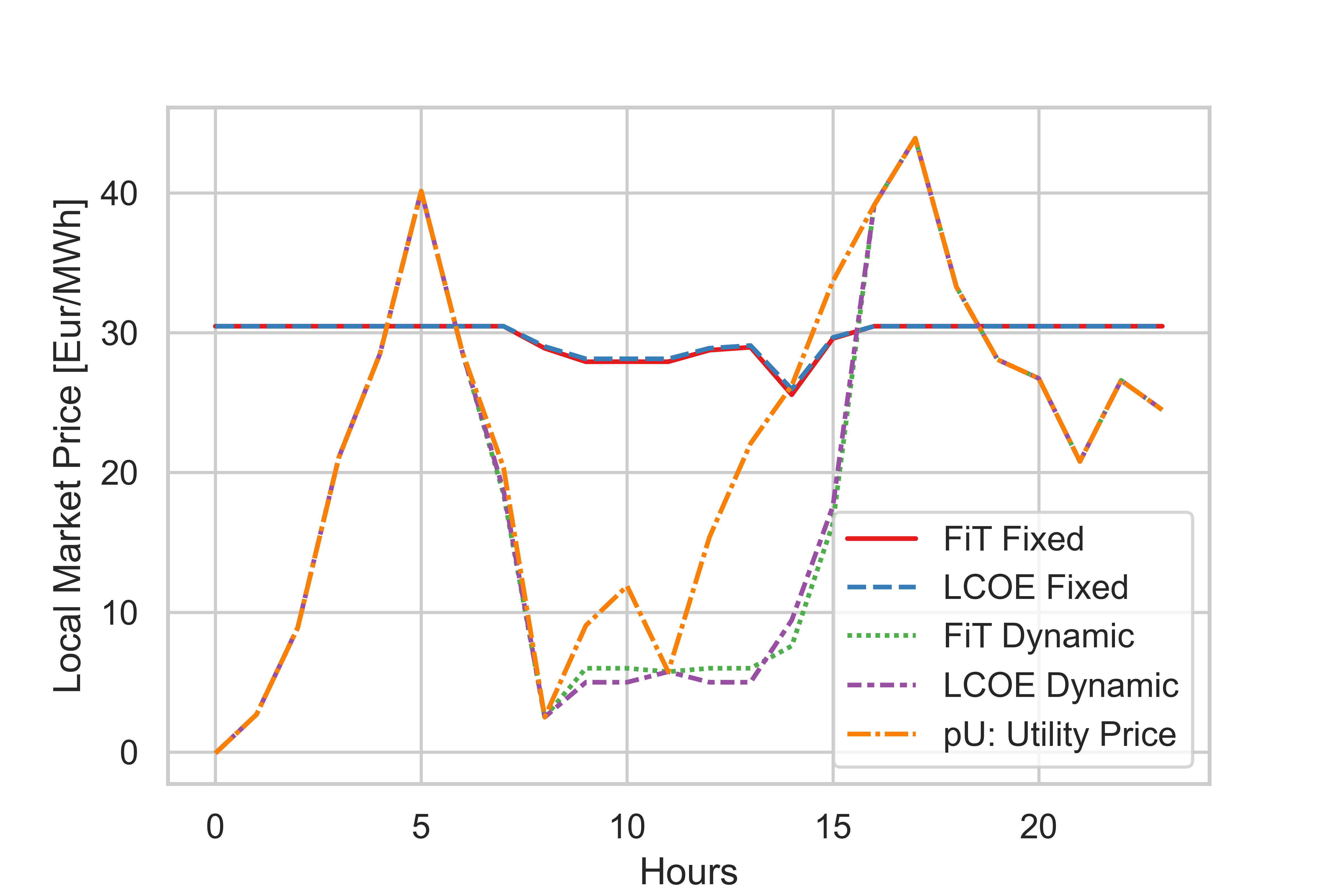}
    \caption{Comparison of \gls{lem} price for a day in March obtained for all the simulated scenarios.}
    \label{fig:Local_price_comp_march}
\end{figure}

Zooming in on the chosen day in March, one can further observe the daily differences between the two pricing schemes, respectively named \textquote{FiT Fixed} and \textquote{LCOE Fixed} in Figure~\ref{fig:Local_price_comp_march}. The fluctuations between the upper and lower bound in the middle of the day when the \gls{lem} is operative are showcased. Even though the price never reaches the lower bound for any of the scenarios for this day, it is still evident that the scheme using the LCOE as the lower bound pushes the \gls{lem} prices down compared to the FiT scheme. 

\begin{figure*}
    \includegraphics[width=\textwidth]{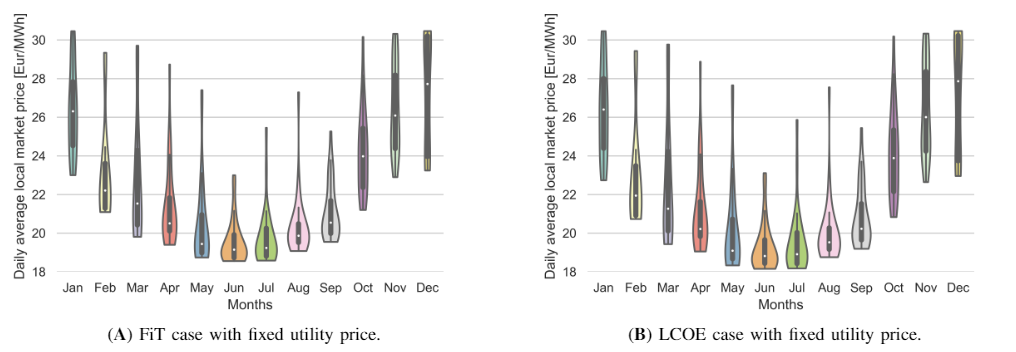}
    \caption{Comparison of the daily average \gls{lem} prices for each month for the (\textbf{A}) FiT scenario and (\textbf{B}) LCOE scenario with fixed utility price.}
    \label{fig:comp_fix_avg}
\end{figure*}

The daily average \gls{lem} prices are shown for each month for the case using the \gls{fit} and fixed pricing thresholds in Figure~\ref{fig:comp_fix_avg}A. The center of gravity of the violins in the plots indicates the frequency of the daily average value for each month, and the white dot represents the monthly average price. As the only \gls{der} within the \gls{lem} is rooftop solar, one can observe that the average prices are lower during summer, due to the amount of surplus in the neighborhood and thus when the $r_t > 1$ condition is occurring. However, the monthly average is lower than the utility price even in the winter months. This means that there is an activity in the \gls{lem} all year, despite the season-dependent generation capacity. Examining the average prices of the LCOE scenario, illustrated in Figure~\ref{fig:comp_fix_avg}B, the \gls{lem} prices are shifted downward, reaching a daily average of around 18 \texteuro/MWh in mid-summer, whereas, the FiT scenario delivers higher minimum \gls{lem} prices of around 19 \texteuro/MWh during summer. The violins are also more stretched in the LCOE scenario, with the highest average at the same value as for the FiT scenario for almost all months, mainly 
due to the same \gls{sdr} effects. It also indicates that the \gls{lem} prices become more volatile using LCOE.

\subsection{Dynamic Pricing Thresholds}
\begin{figure}
    \centering
    \includegraphics[width=3.5in]{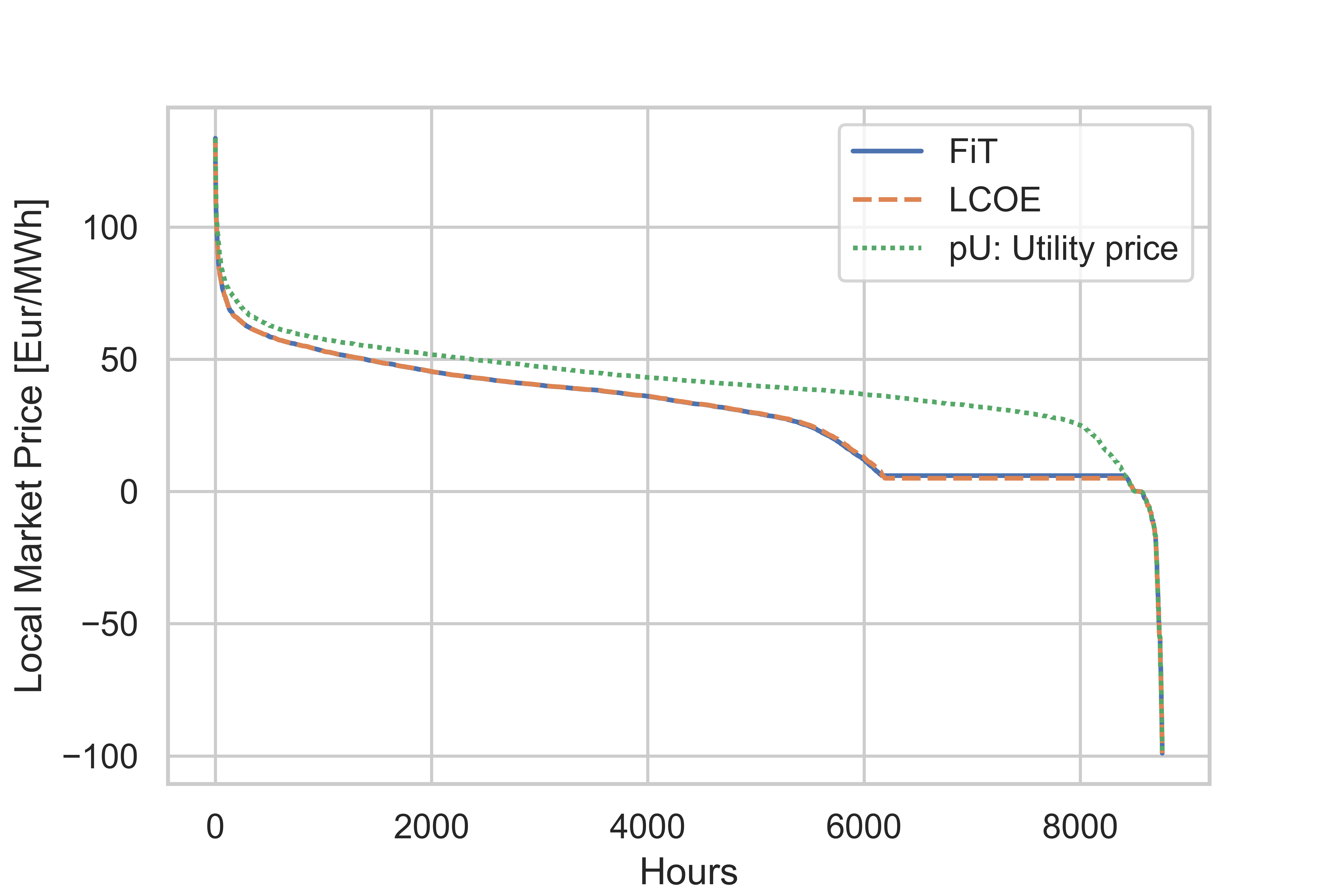}
    \caption{Duration curves of the \gls{lem} prices for the scenarios with dynamic pricing thresholds.}
    \label{fig:Duration_comp_dyn}
\end{figure}

The duration curve for the \gls{lem} price with dynamic pricing thresholds is depicted in Figure~\ref{fig:Duration_comp_dyn}. Similar to the scenarios with fixed pricing thresholds, the LCOE scenario generates slightly lower local prices while the \gls{lem} is active. With the volatile wholesale prices, the \gls{lem} now experiences negative prices. As stated in Section \ref{sec:method_dyn}, these prices only indicate the buying price from the wholesale market, as the prosumers curtail their excess production in these hours.  

Figure~\ref{fig:Local_price_comp_march} shows a snapshot of the \gls{lem} variations of a day in March with the two scenarios with dynamic pricing thresholds, represented by \textquote{FiT Dynamic} and \textquote{LCOE Dynamic}, compared to the supplier price $p_{Ut}$. Both scenarios generate lower prices than the supplier can offer when the \gls{lem} is active. Even though the prices in both scenarios converge to approximately the same level, it is clear that the LCOE scenario mostly obtains the lowest prices. As the auction price, $p_L$, is lower than the FiT value the prices are lower in the LCOE scenario when the market is saturated. When $0<r_{t}<1$, between timestep 13 and 15 in Figure~\ref{fig:Local_price_comp_march}, the prices of the LCOE scenario reaches a slightly higher value than in the FiT scenario, as LCOE$>$FiT.  

\begin{figure*}
    \includegraphics[width=\textwidth]{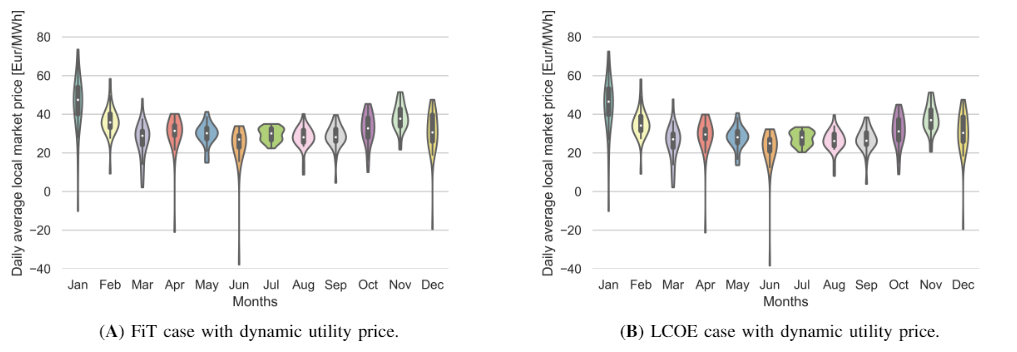}
    \caption{Comparison of the daily average \gls{lem} prices for each month for the (\textbf{A}) FiT scenario and (\textbf{B}) LCOE scenario with dynamic utility price.}
    \label{fig:comp_dyn_avg}
\end{figure*}

The combined effect of LCOE and $p_L$ can be further confirmed by studying Figure~\ref{fig:comp_dyn_avg}, where one can still observe slightly lower values for the LCOE scenario compared to the FiT scenario for all months of the year. Because of the volatile utility price, four days of the year even experience negative daily average prices in both scenarios.  

\subsection{Cost-Benefit Analysis}
\begin{figure*}[h]
    \centering
    \includegraphics[width=\textwidth]{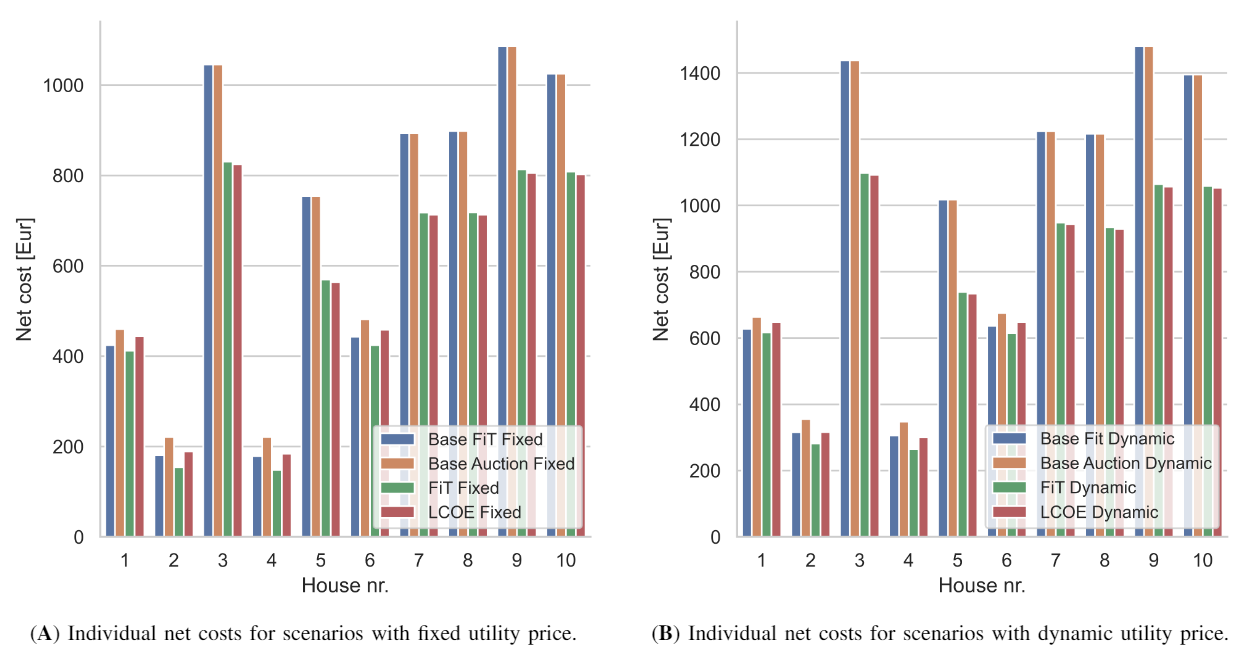}
    \caption{Net costs of all market participants for (\textbf{A}) scenarios with fixed utility price and (\textbf{B}) scenario with dynamic utility price.}
    \label{fig:net_costs}
\end{figure*}

To estimate the full impact of the proposed pricing schemes on both prosumers and consumers participating in the \gls{lem}, a cost-benefit analysis was conducted. The individual net costs of each of the market participants for each of the analyzed scenarios are illustrated in Figure~\ref{fig:net_costs}. Now, the scenarios of the case study are also compared to their respective base scenarios. These are cases where there are no \gls{lem} in place. Hence, the prosumers are only able to sell their surplus to the price offered by the retailer; the FiT, or the auction price, and the consumers must cover their demand through the utility. Thus, for both the fixed and dynamic pricing cases, the net cost of the consumers equals the cost of procuring all energy for the utility price and is the same for both the FiT and the auction alternative. As the auction price is lower than the FiT, the net costs of the prosumers are higher for the \textquote{Base Auction} scenario than for the \textquote{Base FiT} for both the fixed and dynamic cases. 

Looking at the \gls{lem} scenarios, one can observe that the net costs are higher with the LCOE scenarios for all prosumers (house nr. 1, 2, 4, and 6), but lower for all consumers. This is true for both the fixed and dynamic pricing threshold alternatives. Here, it should be noted that the correlation between the $p_L$ and LCOE value, and the utility prices, will depend on the energy mix of the larger electricity market. Although solar power is yet to become a dominant source in the German energy mix, with a share of 8.6\% in 2019 \cite{FraunhoferISERenewableEnergy-Charts}, one can expect that it will become increasingly influential on the electricity price in the future. It can thus be reasoned that if the PV generation is high within the \gls{lem}, it will also be high in the wholesale electricity market. Then, a correlation between the \gls{lem} prices and the utility price is created, such that the utility price will be lower when $r_t$ is high. In this study, historical data for Germany for both pricing and production was used, and the correlation between the two is implicitly considered. 

The overall costs and revenues for the entire neighborhood subject to fixed pricing thresholds are listed in Table~\ref{tab:costs_fix}. Extracting the costs of the consumers, a slight cost reduction is observed, with 0.8\% lower costs than with the FiT scheme. It is also clear that the lower local prices obtained through the LCOE scenario are less beneficial for the prosumers, with a revenue reduction of 13\%. However, comparing the LCOE scheme with the alternative base scenario, where prosumers only get paid the auction value for their surplus, the prosumers' revenue is increased by 14\%. With an additional 22\% reduction in consumer costs, it is clear that using the proposed LCOE pricing scheme through a \gls{lem} setup is much more beneficial for both prosumers and consumers than the originally proposed replacement of the FiT scheme. Lastly, looking at the neighborhood as a whole, the aggregated net costs with fixed pricing are summarized in the last row of Table~\ref{tab:costs_fix}. With the decrease in prosumer revenue in the LCOE scenario, the total net cost is 1.8\% higher than with the FiT scenario, correlating with the individual values in Figure~\ref{fig:net_costs}. Compared to the base scenario with auctions, the total net cost of the neighborhood is reduced by 20\%.

\setlength{\tabcolsep}{1.3pt}
\begin{table*}[h]
    \centering    \caption{Total procurement costs with fixed pricing thresholds, in \texteuro.}
    \begin{tabular}{c|c|c|c|ccc}
        & \textbf{Base FiT} & \textcolor{blue}{\textbf{Base Auction}} & \textcolor{teal}{\textbf{FiT}} & \textbf{LCOE} & \textcolor{teal}{$\Delta $FiT} & \textcolor{blue}{$\Delta$Auction}\\
        \hline
        Total consumer cost ($\sum _{i\in C_{t}} \sum_t c_{it}$) & 5703 & 5703 & 4458 & 4423 & \textcolor{teal}{(-0.8\%)} & \textcolor{blue}{(-22\%)} \\
        Total prosumer revenue ($\sum_t y_t$) & 938 & 782 & 1025 & 888 & \textcolor{teal}{(-13\%)} & \textcolor{blue}{(+14\%)}\\
        Total net cost ($\sum _{i\in C_{t}\cup P_{t}} \sum_t c_{it}^{net}$) & 6929 & 7086 & 5597 & 5699 & \textcolor{teal}{(+1.8\%)} & \textcolor{blue}{(-20\%)}
    \end{tabular}
    \label{tab:costs_fix}
\end{table*}

For the scheme with a dynamic upper pricing threshold, the same effect can be observed for both consumers and prosumers, as demonstrated in Table~\ref{tab:costs_dyn}. The differences between the \gls{lem} scenarios are almost the same as for the fixed pricing threshold cases. The benefits are, however, even more, substantial for the LCOE scheme compared to the base auction scenario. It is important to emphasize that the FiT scenarios in this study are not an alternative as this scheme is being phased out, and is solely used to benchmark the new, proposed pricing mechanism. As auctions are one of the solutions being suggested as a replacement of the FiT scheme by German policymakers, the comparison to this scenario is much more relevant.

\begin{table*}[h]
    \centering
    \caption{Total procurement costs with dynamic pricing thresholds, in \texteuro.}
    \begin{tabular}{c|c|c|c|ccc}
        & \textbf{Base FiT} & \textcolor{blue}{\textbf{Base Auction}} & \textcolor{teal}{\textbf{FiT}} & \textbf{LCOE} & \textcolor{teal}{$\Delta $FiT} & \textcolor{blue}{$\Delta$Auction}\\
        \hline
        Total consumer cost ($\sum _{i\in C_{t}} \sum_t c_{it}$) & 7771 & 7771 & 5843 & 5809 & \textcolor{teal}{(-0.6\%)} & \textcolor{blue}{(-25\%)} \\
        Total prosumer revenue ($\sum_t y_t$) & 938 & 782 & 1044 & 909 & \textcolor{teal}{(-13\%)} & \textcolor{blue}{(+16\%)}\\
        Total net cost ($\sum _{i\in C_{t}\cup P_{t}} \sum_t c_{it}^{net}$) & 9657 & 9813 & 7621 & 7722 & \textcolor{teal}{(+1.3\%)} & \textcolor{blue}{(-21\%)}
    \end{tabular}
    \label{tab:costs_dyn}
\end{table*}

\subsection{Sensitivity Analysis}
To analyze the actual impact of the \gls{sdr} on the \gls{lem} price, a sensitivity analysis was conducted by varying the amount of surplus in the neighborhood. Additional simulations were thus performed, increasing the number of prosumers for each time. 

\begin{figure}[h]
    \centering
    \includegraphics[width=\linewidth]{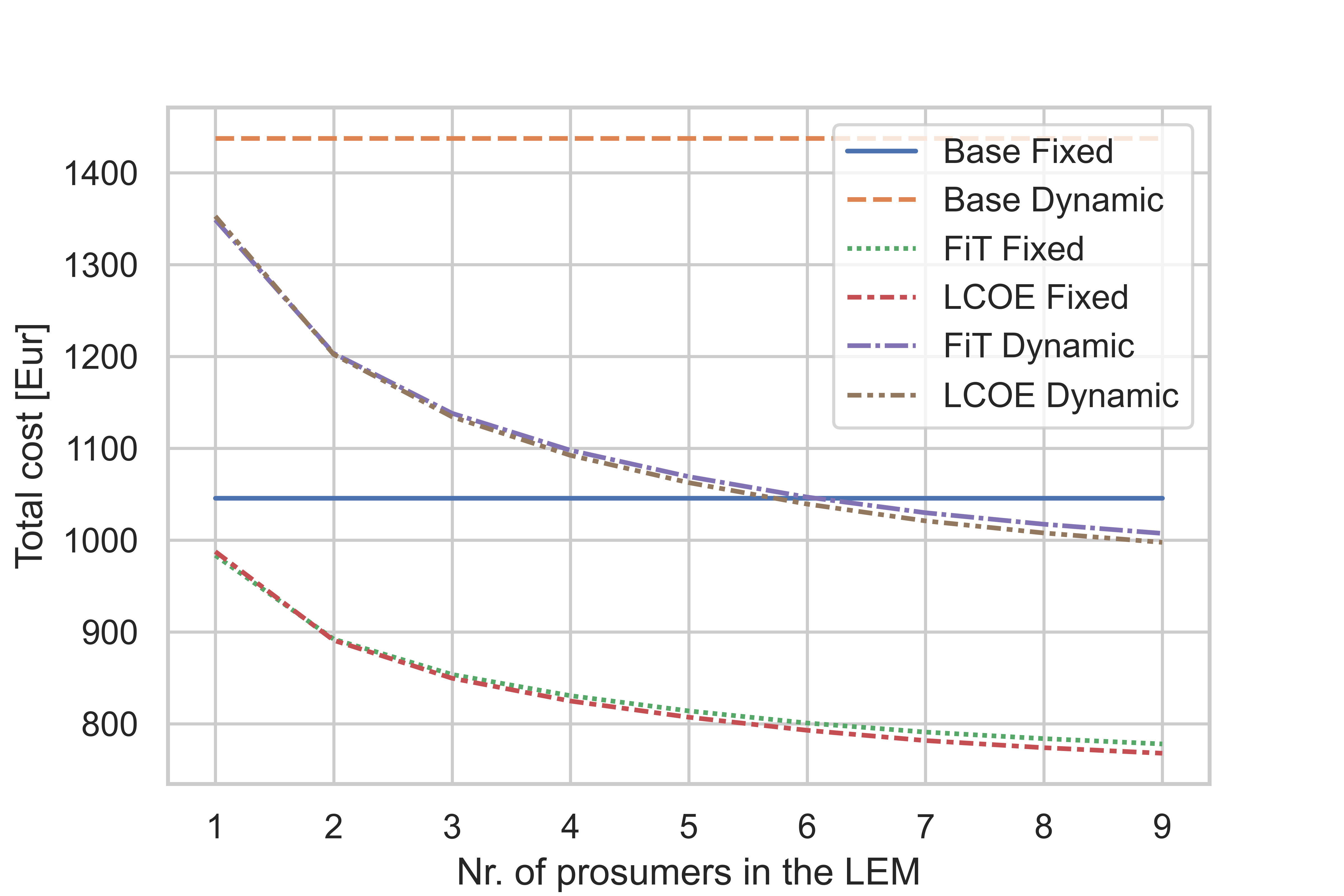}
    \caption{Annual cost for Consumer 3, depending on number of prosumers in the neighborhood.}
    \label{fig:sens_consumer}
\end{figure}

House 3 retains the role of a pure consumer for all simulations. Its net costs, depending on the number of prosumers in the neighborhood, are plotted in Figure~\ref{fig:sens_consumer}. The alternatives of procuring everything for both the fixed and dynamic utility prices are also plotted for comparison. In general, the consumer costs decrease as more surplus becomes available in the \gls{lem}. For both schemes based on fixed and dynamic pricing thresholds, the LCOE scenario generates the lowest net costs for the consumers, independent of the market saturation. The difference between the LCOE and FiT scenario increase along with the increasing surplus.

\begin{figure}[h]
    \centering
    \includegraphics[width=3.5in]{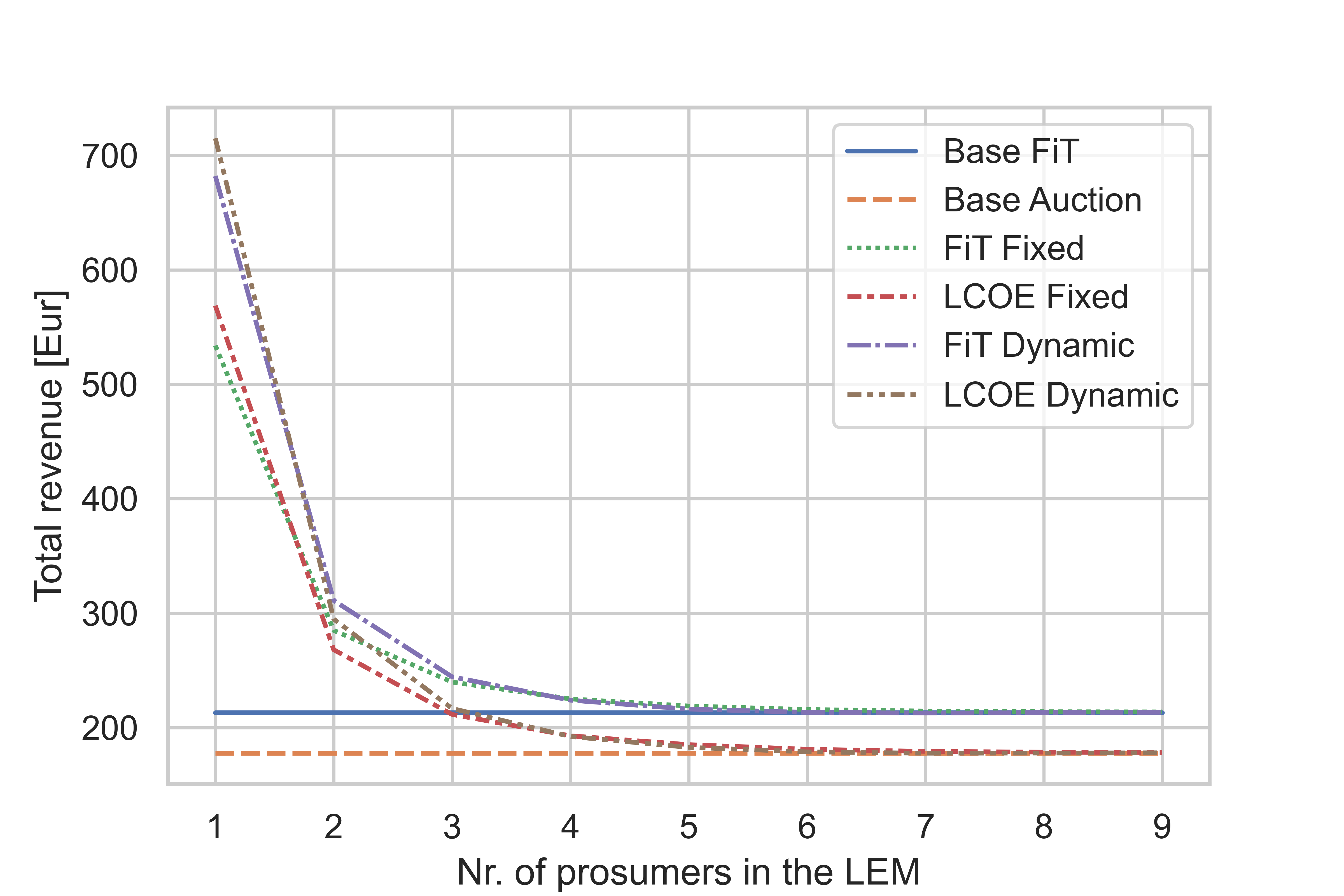}
    \caption{Annual revenue for Prosumer 1, depending on the number of prosumers in the neighborhood.}
    \label{fig:sens_prosumer}
\end{figure}

House 1 acts as a prosumer in all simulated cases with increasing numbers of prosumers in the \gls{lem}. Its total revenue is depicted in Figure~\ref{fig:sens_prosumer}. As the electricity consumption of the neighborhood remains the same for all cases, one would expect that each prosumer's revenue would decrease when more surplus becomes available in the market. Accordingly, it is clear from Figure~\ref{fig:sens_prosumer}, that the prosumer revenue in all the LEM cases converges towards the base scenarios where all surplus is sold for the price offered by the utility. With six or more prosumers in this particular neighborhood, the market reaches its saturation point; in other words, $r_t>1$ frequently with more than three prosumers. This means that the demand of the neighborhood is covered, and the prosumers often sell the remaining surplus to the utility. As both the FiT and $p_L$ value used in this study is lower than the LCOE, the viability of the prosumers' DER investments is then hard to maintain. Thus, the incentives of becoming a prosumer under these \gls{lem} conditions are diminishing. With the additional decreasing cost for the consumers, however, one can also assume that it becomes more attractive to join the \gls{lem} as a consumer, stabilizing the supply-demand balance in the neighborhood.

In Germany, it is estimated that a regular household owning a PV system, consumes 20-40\% of their self-generated electricity \cite{WirthRecentGermany}. In all simulated scenarios, around 52\% of the PV power was sold to the utility, regardless of the \gls{lem} pricing scheme. This indicates that with a \gls{lem} in place, more locally generated electricity remains within the neighborhood. Thus, in terms of increasing self-consumption and activating the \gls{lem}, prosumers would benefit from installing batteries. This could also postpone the saturation point observed in Figure~\ref{fig:sens_prosumer}, as the surplus available in the \gls{lem} can be more evenly distributed across the day. Then, the attractiveness of investing in DER equipment with several prosumers already in the market is restored. As the pricing scheme replacing the FiT is yet to be determined it is also worth noting that the price offered by the utility to buy surplus from the prosumers can be affected by the availability of storage. With a high amount of unpredictable PV power being delivered from the \gls{lem}, the utility may end up having to settle a lot in the balancing and intraday market, and may thus be less willing to offer a good price for the surplus. Predictability and flexibility can then be of higher value.

\section{Conclusion}
To promote \glspl{lem} as an attractive option to take an active role in the energy system for end-users, a transparent and fair pricing mechanism for locally traded energy must be established. With the phase-out of the FiT and similar support schemes, other methods to ensure the profitability of participating in local flexibility activities must be redesigned. The average LCOE value of the \gls{lem} is proposed in this paper, showing promising results in terms of cost reductions for the consumers in the market. Comparing the new LCOE convention with the existing FiT scheme, the prosumers' revenue is decreased. However, comparing it to the auction alternative, both consumers and prosumers experience substantially higher benefits. This finding can also be utilized by policymakers to incentivize the decarbonization, decentralization, and democratization through LEMs in areas with a high penetration level of DERs, and thus further contribute to UN SDG \#7. Hence, the proposed pricing mechanism requires sophistication, with track recording of parameters like the SDR, $r_t$, dynamically, while also ensuring information security for the participating agents. DLT is thus highlighted as one of the most promising digitalization technologies to provide a functional solution to realize such next-generation energy policy and LEM mechanisms. Still, the proposed pricing scheme is easy to grasp for its users, and will likely be perceived as fair by the market participants, as the costs and benefits are distributed according to actual contribution to the market. Further improvements of the method should involve the inclusion of energy storage technology and how it influences both the LCOE value and the strategic behavior of the market agents.        

\section*{Acknowledgment}
We are grateful for the financial support granted by the Norwegian Research Council through the IDUN project, under grant agreement No 295920.

\bibliographystyle{IEEEtran}
\bibliography{references}

\begin{thebibliography}{10}
\providecommand{\url}[1]{#1}
\csname url@samestyle\endcsname
\providecommand{\newblock}{\relax}
\providecommand{\bibinfo}[2]{#2}
\providecommand{\BIBentrySTDinterwordspacing}{\spaceskip=0pt\relax}
\providecommand{\BIBentryALTinterwordstretchfactor}{4}
\providecommand{\BIBentryALTinterwordspacing}{\spaceskip=\fontdimen2\font plus
\BIBentryALTinterwordstretchfactor\fontdimen3\font minus
  \fontdimen4\font\relax}
\providecommand{\BIBforeignlanguage}[2]{{%
\expandafter\ifx\csname l@#1\endcsname\relax
\typeout{** WARNING: IEEEtran.bst: No hyphenation pattern has been}%
\typeout{** loaded for the language `#1'. Using the pattern for}%
\typeout{** the default language instead.}%
\else
\language=\csname l@#1\endcsname
\fi
#2}}
\providecommand{\BIBdecl}{\relax}
\BIBdecl

\bibitem{Tushar2021Peer-to-peerChallenges}
W.~Tushar, C.~Yuen, T.~K. Saha, T.~Morstyn, A.~C. Chapman, M.~J.~E. Alam,
  S.~Hanif, and H.~V. Poor, ``{Peer-to-peer energy systems for connected
  communities: A review of recent advances and emerging challenges},''
  \emph{Applied Energy}, vol. 282, p. 116131, 1 2021.

\bibitem{Bjarghov2021DevelopmentsReview}
S.~Bjarghov, M.~Loschenbrand, A.~U. Ibn~Saif, R.~Alonso~Pedrero, C.~Pfeiffer,
  S.~K. Khadem, M.~Rabelhofer, F.~Revheim, and H.~Farahmand, ``{Developments
  and Challenges in Local Electricity Markets: A Comprehensive Review},''
  \emph{IEEE Access}, vol.~9, pp. 58\,910--58\,943, 2021.

\bibitem{Hahnel2020BecomingCommunities}
U.~J. Hahnel, M.~Herberz, A.~Pena-Bello, D.~Parra, and T.~Brosch, ``{Becoming
  prosumer: Revealing trading preferences and decision-making strategies in
  peer-to-peer energy communities},'' \emph{Energy Policy}, vol. 137, 2 2020.

\bibitem{Lin2019ComparativeMarkets}
J.~Lin, M.~Pipattanasomporn, and S.~Rahman, ``{Comparative analysis of auction
  mechanisms and bidding strategies for P2P solar transactive energy
  markets},'' \emph{Applied Energy}, vol. 255, no. August, p. 113687, 2019.

\bibitem{Ghorani2018OptimalMarkets}
R.~Ghorani, M.~Fotuhi-Firuzabad, and M.~Moeini-Aghtaie, ``{Optimal Bidding
  Strategy of Transactive Agents in Local Energy Markets},'' \emph{IEEE
  Transactions on Smart Grid}, 2018.

\bibitem{Wang2020ASystem}
Z.~Wang, X.~Yu, Y.~Mu, and H.~Jia, ``{A distributed Peer-to-Peer energy
  transaction method for diversified prosumers in Urban Community Microgrid
  System},'' \emph{Applied Energy}, vol. 260, no.~92, p. 114327, 2020.

\bibitem{Vieira2021Peer-to-peerContracts}
G.~Vieira and J.~Zhang, ``{Peer-to-peer energy trading in a microgrid leveraged
  by smart contracts},'' \emph{Renewable and Sustainable Energy Reviews}, vol.
  143, p. 110900, 6 2021.

\bibitem{Morstyn2020IntegratingPricing}
T.~Morstyn, A.~Teytelboym, C.~Hepburn, and M.~D. McCulloch, ``{Integrating P2P
  Energy Trading with Probabilistic Distribution Locational Marginal
  Pricing},'' \emph{IEEE Transactions on Smart Grid}, vol.~11, no.~4, pp.
  3095--3106, 2020.

\bibitem{Paudel2019PricingApproach}
A.~Paudel and H.~B. Gooi, ``{Pricing in Peer-to-Peer Energy Trading Using
  Distributed Optimization Approach},'' \emph{IEEE Power and Energy Society
  General Meeting}, vol. 2019-August, pp. 8--12, 2019.

\bibitem{Ullah2020DistributedGrid}
M.~H. Ullah, A.~Alseyat, and J.~D. Park, ``{Distributed Dynamic Pricing in
  Peer-to-Peer Transactive Energy Systems in Smart Grid},'' \emph{IEEE Power
  and Energy Society General Meeting}, vol. 2020-Augus, pp. 1--5, 2020.

\bibitem{Morstyn2019MulticlassPreferences}
T.~Morstyn and M.~D. McCulloch, ``{Multiclass Energy Management for
  Peer-to-Peer Energy Trading Driven by Prosumer Preferences},'' \emph{IEEE
  Transactions on Power Systems}, vol.~34, no.~5, pp. 4005--4014, 2019.

\bibitem{Lezama2020LearningOptimization}
F.~Lezama, R.~Faia, J.~Soares, P.~Faria, and Z.~Vale, ``{Learning Bidding
  Strategies in Local Electricity Markets using Ant Colony optimization},''
  \emph{2020 IEEE Congress on Evolutionary Computation, CEC 2020 - Conference
  Proceedings}, 7 2020.

\bibitem{Jiang2020ElectricityEnvironment}
Y.~Jiang, K.~Zhou, X.~Lu, and S.~Yang, ``{Electricity trading pricing among
  prosumers with game theory-based model in energy blockchain environment},''
  \emph{Applied Energy}, vol. 271, p. 115239, 8 2020.

\bibitem{Fernandez2021APeers}
E.~Fernandez, M.~J. Hossain, K.~Mahmud, M.~S.~H. Nizami, and M.~Kashif, ``{A
  Bi-level optimization-based community energy management system for optimal
  energy sharing and trading among peers},'' \emph{Journal of Cleaner
  Production}, vol. 279, p. 123254, 1 2021.

\bibitem{Cali2019EnergyMarkets}
U.~Cali and O.~Cakir, ``{Energy Policy Instruments for Distributed Ledger
  Technology Empowered Peer-to-Peer Local Energy Markets},'' \emph{IEEE
  Access}, vol.~7, pp. 82\,888--82\,900, 2019.

\bibitem{Grzanic2021ElectricityUncertainty}
M.~Grzanic, J.~M. Morales, S.~Pineda, and T.~Capuder, ``{Electricity
  Cost-Sharing in Energy Communities under Dynamic Pricing and Uncertainty},''
  \emph{IEEE Access}, vol.~9, pp. 30\,225--30\,241, 2021.

\bibitem{Long2018Peer-to-peerMicrogrid}
C.~Long, J.~Wu, Y.~Zhou, and N.~Jenkins, ``{Peer-to-peer energy sharing through
  a two-stage aggregated battery control in a community Microgrid},''
  \emph{Applied Energy}, vol. 226, pp. 261--276, 9 2018.

\bibitem{Anoh2020}
K.~Anoh, S.~Maharjan, A.~Ikpehai, Y.~Zhang, and B.~Adebisi, ``Energy
  peer-to-peer trading in virtual microgrids in smart grids: A game-theoretic
  approach,'' \emph{IEEE Transactions on Smart Grid}, vol.~11, no.~2, pp.
  1264--1275, 2020.

\bibitem{Tushar2017PriceApproach}
W.~Tushar, C.~Yuen, D.~B. Smith, and H.~V. Poor, ``{Price Discrimination for
  Energy Trading in Smart Grid: A Game Theoretic Approach},'' \emph{IEEE
  Transactions on Smart Grid}, vol.~8, no.~4, pp. 1790--1801, 2017.

\bibitem{Liu2017Energy-SharingProsumers}
N.~Liu, X.~Yu, C.~Wang, C.~Li, L.~Ma, and J.~Lei, ``{Energy-Sharing Model with
  Price-Based Demand Response for Microgrids of Peer-to-Peer Prosumers},''
  \emph{IEEE Transactions on Power Systems}, vol.~32, no.~5, pp. 3569--3583, 9
  2017.

\bibitem{Zhou2018EvaluationFramework}
Y.~Zhou, J.~Wu, and C.~Long, ``{Evaluation of peer-to-peer energy sharing
  mechanisms based on a multiagent simulation framework},'' \emph{Applied
  Energy}, vol. 222, pp. 993--1022, 7 2018.

\bibitem{An2020DeterminingMicrogrid}
J.~An, M.~Lee, S.~Yeom, and T.~Hong, ``{Determining the Peer-to-Peer
  electricity trading price and strategy for energy prosumers and consumers
  within a microgrid},'' \emph{Applied Energy}, vol. 261, p. 114335, 3 2020.

\bibitem{Andoni2019BlockchainOpportunities}
M.~Andoni, V.~Robu, D.~Flynn, S.~Abram, D.~Geach, D.~Jenkins, P.~McCallum, and
  A.~Peacock, ``{Blockchain technology in the energy sector: A systematic
  review of challenges and opportunities},'' \emph{Renewable and Sustainable
  Energy Reviews}, vol. 100, pp. 143--174, 2 2019.

\bibitem{Khorasany2021LightweightTrading}
\BIBentryALTinterwordspacing
M.~Khorasany, A.~Dorri, R.~Razzaghi, and R.~Jurdak, ``{Lightweight blockchain
  framework for location-aware peer-to-peer energy trading},''
  \emph{International Journal of Electrical Power and Energy Systems}, vol.
  127, no. October 2020, p. 106610, 2021. [Online]. Available:
  \url{https://doi.org/10.1016/j.ijepes.2020.106610}
\BIBentrySTDinterwordspacing

\bibitem{Cali2021NovelProblem}
U.~Cali and O.~Cakir, ``{Novel donation sharing mechanisms under smart energy
  cyber-physical-social system and DLT to contend the energy poverty
  problem},'' \emph{IEEE Access}, vol.~9, pp. 127\,037--127\,053, 2021.

\bibitem{Vartiainen2020}
E.~Vartiainen, G.~Masson, C.~Breyer, D.~Moser, and E.~Rom{\'{a}}n~Medina,
  ``{Impact of weighted average cost of capital, capital expenditure, and other
  parameters on future utility-scale PV levelised cost of electricity},''
  \emph{Progress in Photovoltaics: Research and Applications}, vol.~28, no.~6,
  pp. 439--453, 2020.

\bibitem{Hoogsteen2017AManagement}
G.~Hoogsteen, ``{A Cyber-Physical Systems Perspective on Decentralized Energy
  Management},'' Ph.D. dissertation, University of Twente, 2017.

\bibitem{Pfenninger2016Long-termData}
S.~Pfenninger and I.~Staffell, ``{Long-term patterns of European PV output
  using 30 years of validated hourly reanalysis and satellite data},''
  \emph{Energy}, vol. 114, pp. 1251--1265, 11 2016.

\bibitem{Rienecker2011MERRA:Applications}
\BIBentryALTinterwordspacing
M.~M. Rienecker, M.~J. Suarez, R.~Gelaro, R.~Todling, J.~Bacmeister, E.~Liu,
  M.~G. Bosilovich, S.~D. Schubert, L.~Takacs, G.~K. Kim, S.~Bloom, J.~Chen,
  D.~Collins, A.~Conaty, A.~Da~Silva, W.~Gu, J.~Joiner, R.~D. Koster,
  R.~Lucchesi, A.~Molod, T.~Owens, S.~Pawson, P.~Pegion, C.~R. Redder,
  R.~Reichle, F.~R. Robertson, A.~G. Ruddick, M.~Sienkiewicz, and J.~Woollen,
  ``{MERRA: NASA’s Modern-Era Retrospective Analysis for Research and
  Applications},'' \emph{Journal of Climate}, vol.~24, no.~14, pp. 3624--3648,
  7 2011. [Online]. Available:
  \url{https://journals.ametsoc.org/view/journals/clim/24/14/jcli-d-11-00015.1.xml}
\BIBentrySTDinterwordspacing

\bibitem{PriceGermany}
{BDEW}, ``{BDEW-Strompreisanalyse Januar 2022},''
  \url"https://www.bdew.de/service/daten-und-grafiken/bdew-strompreisanalyse/",
  1 2022, [Date Accessed: 2022-2-25].

\bibitem{FraunhoferISESpotEnergy-Charts}
\BIBentryALTinterwordspacing
{Fraunhofer ISE}, ``{Spot Market Prices - Energy-Charts},'' 2022, [Date
  Accessed: 2022-2-7]. [Online]. Available:
  \url{https://energy-charts.info/charts/price\_spot\_market/chart.htm?l=en\&c=DE\&year=2019\&stacking=stacked_absolute_area\&interval=year\&legendItems=000001}
\BIBentrySTDinterwordspacing

\bibitem{Renews}
C.~Morris, ``Community energy in germany - more than just climate change
  mitigation,'' \emph{Renewable Energy Agency}, no.~89, 2019.

\bibitem{Halden2021DLT-basedInvestments}
U.~Halden, U.~Cali, M.~F. Dynge, J.~Stekli, and L.~Bai, ``{DLT-based equity
  crowdfunding on the techno-economic feasibility of solar energy
  investments},'' \emph{Solar Energy}, vol. 227, pp. 137--150, 10 2021.

\bibitem{Kost2021}
C.~Kost, S.~Shammugam, V.~Fluri, D.~Peper, A.~D. Memar, and T.~Schlegl,
  ``{Levelized Cost of Electricity - Renewable Energy Technologies, June
  2021},'' \emph{Fraunhofer Institute for Solar Energy Systems ISE}, no. June,
  2021.

\bibitem{FraunhoferISERenewableEnergy-Charts}
\BIBentryALTinterwordspacing
{Fraunhofer ISE}, ``{Renewable Shares - Energy-Charts},'' 2022, [Date Accessed:
  2022-2-7]. [Online]. Available:
  \url{https://energy-charts.info/charts/renewable\_share/chart.htm?l=en\&c=DE\&share=solar\_share\&interval=year}
\BIBentrySTDinterwordspacing

\bibitem{WirthRecentGermany}
\BIBentryALTinterwordspacing
H.~Wirth, ``{Recent Facts about Photovoltaics in Germany},'' 2021, version of
  May 15, 2021. [Online]. Available:
  \url{https://www.ise.fraunhofer.de/content/dam/ise/en/documents/publications/studies/recent-facts-about-photovoltaics-in-germany.pdf}
\BIBentrySTDinterwordspacing

\end{thebibliography}


\end{document}